# Mobility profiles and calendars for food security and livelihoods analysis

Pedro J. Zufiria [1], David Pastor-Escuredo [1], Luis Úbeda-Medina [1], Miguel A. Hernández-Medina [1], Iker Barriales-Valbuena [1], Alfredo J. Morales [1], Wilfred Nkwambi [2], John Quinn [3], Paula Hidalgo-Sanchís [3], Miguel Luengo-Oroz [3]

1 Universidad Politécnica de Madrid
2 United Nations World Food Program Senegal
3 Pulse Lab Kampala, United Nations Global Pulse

## 1. INTRODUCTION

Social vulnerability is defined as "the capacity of individuals and social groups to respond to any external stress placed on their livelihoods and well-being" [4]. Mobility and migrations are relevant when assessing vulnerability since the movements of a population reflect on their livelihoods, coping strategies and social safety nets. Although in general migration characterization is complex and open to controversy [6], changes in mobility patterns for vulnerable population groups are likely to indicate a change in livelihoods or coping strategies. These changes can also indicate that the population groups may be exposed to new shocks; hence, monitoring of changes in mobility patterns can be a powerful early warning mechanism.

Livelihoods in Senegal show a strong correlation with geographical location, and have been mapped out for analysis in different zones such as pastoralism, agriculture and fishing [2]. Within each of these zones, there are well-studied patterns of seasonal activities and population movements. However, such changes have until now been impossible to observe directly. Telecoms data therefore provide an important new opportunity to observe such changes in mobility patterns in real-time. For this purpose, we have developed statistical measures for profiling and calendarizing mobility in the context of livelihood zones and seasonal activity patterns in Senegal.

For each of the 13 mapped Livelihood Zones (LZ) of Senegal, we have characterized the profiles and calendars of the mobility flows from/to other LZs with different livelihood conditions. We have classified the population according to their mobility behaviors by clustering individual mobility trajectories into mobility classes. The timing of the displacements for each of the "mobility classes" has been aligned and compared with seasonal calendars and rainfall information. The calendar framework can be used to generate mobility baselines that combined with future real time data access could contribute food security early warning mechanisms.

## 2. MATERIALS AND METHODS

The proposed analysis is based on a model which gathers all the different types of variables considered to be relevant for characterizing any user mobility behavior. The model helps to integrate and analyse heterogeneous data with different time and space resolutions, by adjusting the domain of the variables from days to months or from antennas to livelihoods.

We start presenting the basic model variables based on the available data from the D4D datasets and external resources, newly defined variables and the developed analysis procedures.

### 2.1. Modelling variables

Here we present the different types of variables and their relationship with the available data.

#### 2.1.1. Basic variables

The variables characterizing telephone users can be classified into:

1. User Behavior variables, *UB(t) = (l(t); c(t))*, gather both his/her geographical location *l = ($l_a$, $l_o$)*, and communication status *c* along time.
2. Environment variables, *E($l_e$, t)*, affect user behavior and depend on geographical location $l_e$ and time *t* (e.g., rainfalls, holidays, etc.).
3. Indicators, *I($l_i$, t)*, gather other relevant variables one may want to characterize (e.g., level of food insecurity of location $l_i$ at time *t*, etc.).

#### 2.1.2. Derived secondary variables

Secondary variables can be derived from the basic variables. We can define two types:
1. User derived variables group the information (via time and/or space aggregation), keeping the (anonymized) user ID label. There are two types:
    a. Variables for which data are available (see Section 2.1.2.1).
    b. Variables defined for methodological purposes (see Section 2.1.2.2).
2. Environment or Indicator derived variables (see also Section 2.1.2.3).

#### 2.1.2.1. Available data variables

The variables for which data are available in this challenge are:

- user bandicoot indicators *b(t)*. Both Data-set 2 and 3, provide measurements of these variables in a monthly averaged basis.
- user Arrondissement location *A(t) = A(l(t))* (derived from *l(t)*). Data-set 3 provides measurements of this variable for each user along the whole year.

2.1.2.2. Method related variables: Home or preferential location

These variables are required for the proposed methodology. They are relevant latent variables, since most environment variables and indicators depend both on space and time. Such variables can be estimated with different time and geographical resolutions, depending on the data employed. Based on Data-set 3, time aggregation procedures provide estimations of Daily-Home Arrondissement (DHA) and Monthly-Home Arrondissement (MHA) for each user. They can be complemented with the geographical location of the centroids corresponding to each Arrondissement. In addition, a geographical aggregation allows to consider the Monthly-Home Livelihood Zones (MHLZ) for each user. The D4D contextual data (shapefiles) have been used to aggregate the population from BTS to Arrondissements and from Arrondissements to regions or Livelihoods Zones (Figures 1a and 1b illustrate different levels of geographical resolution).

2.1.2.3. Daily Rainfall (DR) by Arrondissements or Livelihood Zones

They are obtained from a geographical aggregation of NASA's TRMM sensed data [9], collected with a 0.25 resolution (longitude and latitude) in a daily basis.

**2.2. Defining user feature vectors**

MHA and MHLZ provide the location of users over time for the whole year with an Arrondissement and Livelihood Zone-month resolution respectively; this is complemented with the bandicoot information provided in Data-set 3, to define the feature vectors:

- Home Arrondissement User Vector (HAUV): A 13-dimensional vector comprising the user ID and his/her MHA for the 12 months.
- Home Livelihood Zone User Vector (HLZUV): A 13-dimensional vector comprising the user ID and his/her MHLZ for the 12 months (according to the map of Fig.1a/b)
- Bandicoot User Vectors (BUVs): for each bandicoot we have a 13-dimensional vector comprising the user ID and of his/her bandicoot value for the 12 months.

Our main objective is to unravel mobility patterns from the analysis of the HAUV, HLZUV and BUVs together with DR. Such analysis is aimed to classify the population mobility behavior into different groups depending on the period of the year and their geographical location. The results are complemented with the detection of general population movements associated with relevant events.

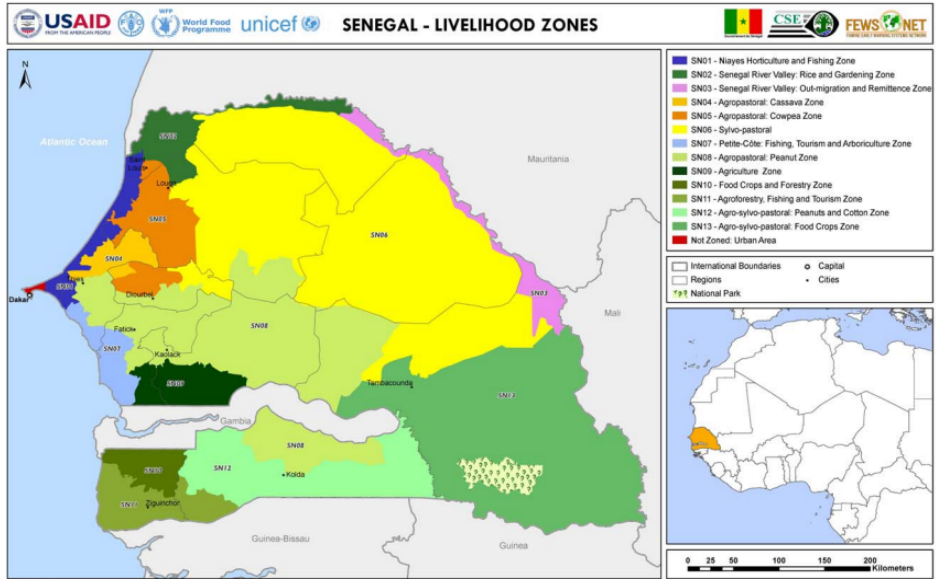

Fig. 1a: Livelihood zones map in Senegal. This map has been used to generate an Arrondissement to Livelihood assignment.

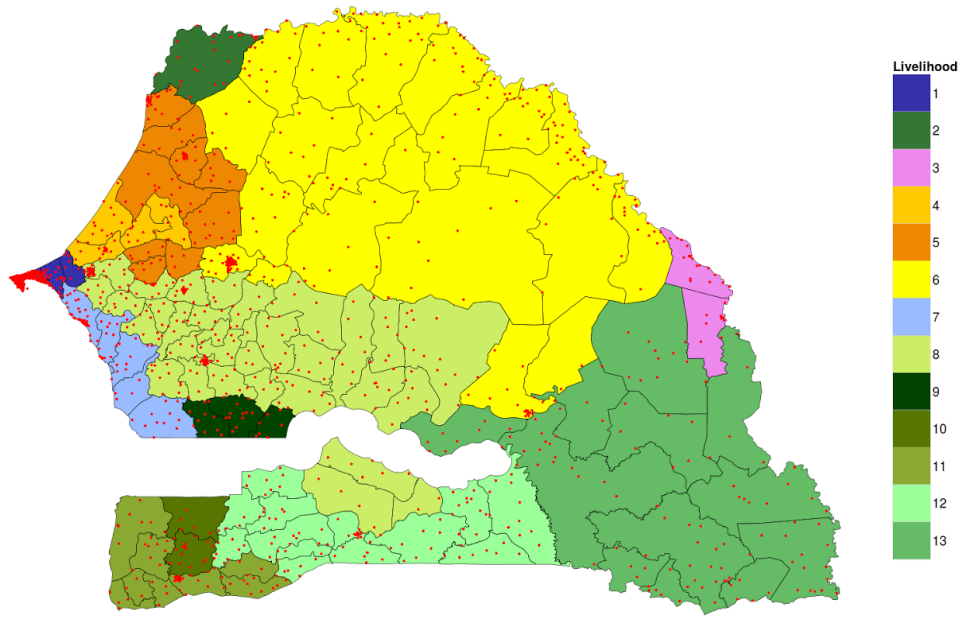

Fig. 1b: Levels of geographical resolution in Senegal based upon D4D datasets. The coverage regions of the antennas (red dots) correspond to Data-set 2 and can be aggregated to the Arrondissement level (line boundaries). In this work, the Arrondissements (Data-set 3) are grouped by Livelihood Zones (colors) rather than political boundaries.

## 2.3. Classification of feature vectors

### 2.3.1. Pre-filtering of mobility profiles

When considering mobility profiles, users can be filtered depending on different criteria. For instance, users whose HAUV or HLZUV components are all equal can be removed (considered as "non-moving" users). In addition, users for which the geographical distance corresponding to their Arrondissement change does not surpass a given ratio with respect to their radius of gyration (obtained from bandicoot data), can be also removed as "regular-travelers" and so forth. These filtering criteria were tested before the classification procedures (to be explained in the following Section 2.3.2).

### 2.3.2. Population clustering

Since a global country classification analysis does not seem feasible and useful for assessing livelihood related mobility patterns, the analysis has been performed by regions (at the Arrondissement and Livelihood Zone levels). At the Arrondissement level several alternatives have been evaluated for classification: if Arrondissement centroids or numbers are considered, the results, though very rich in terms of detailed information, are not easy to interpret. A binary representation indicating if the user is or not in the Arrondissement under consideration seems more tractable but the overall geographical interpretation remains complex. Therefore, the final analysis was addressed at a Livelihood Zone (LZ) level.

For each LZ, users who have visited it for some period of the year have been considered together, and their HLZUV analyzed. HLZUVs can be classified into different groups attending to mobility profiles. The binary representation indicating if the user is or not in the Livelihood Zone under consideration has been selected since it is easy to deal with and allows simple interpretations.

A new stage filtering procedure can be performed on the binarized HLZUV, depending upon several parametric temporal consistency constraints to remove noisy trajectories that may appear as a result of singular mobility profile or inaccuracy in the home location estimation:

- The user must have stayed at least $M_{min}$ consecutive months in the target LZ.
- The user must have not stayed more than $M_{max}$ months in the target LZ.
- The user must have stayed at least $M_{outmin}$ months in some other LZ.
- The user must have stayed at a specific period of the year (when looking for specific types of mobility profiles, such as the ones related to rainfalls).

Then, the remaining binarized HLZUV are grouped into classes. Such clustering can be performed in different ways depending on:

- The type distance defined between vectors. Ultimately, the metric used should reflect the specific perspective of users' behavior similarity in terms of mobility profile. So far several distances (Euclidean, Manhattan, Cosine) have been used to generate distance distribution between vector pairs, and they seem to provide similar results.
- The clustering procedure. Hierarchical clustering has been employed using a grouping method relying on the "average" distance to build the tree nodes. The provided dendrogram tree is cut by a maximum number of representative classes that may vary between 4-5 classes for each LZ. Each of the cluster classes stands for a mobility profile class within the population that has occupied the target LZ under the constraints imposed.

The trajectories grouped together in different clusters, provide typical consistent mobility profiles that can be used as seed information to understand migrations and social behaviours to seasonal changes or large scale events.

## 2.4. Time gradients for event detection and period selection

The computation of time differences (or gradients) together with a threshold-based detection scheme have been employed for global event detection and for relevant period selection.

### 2.4.1. Global event detection

Global event detection has been performed by analysing aggregated HAUV vectors; when aggregating (by users) this information, general population movement behaviors can be detected which are associated with relevant events in the country.

### 2.4.2. Time period selection for cluster analysis

Similar gradient computations on the profiles associated with user in a cluster provides the relevant periods of time where most movements occur, allowing for a more specific analysis.

## 2.5. Complementary processing

### 2.5.1. Class characterization based on monthly locations and bandicoot data

Using the user IDs of each class, the corresponding bandicoot vectors have been classed up together and statistically characterized with the mean and std, obtaining a behavioral characterization of each class.

At the Arrondissement resolution level, a "Distance to Home Vector" (DHV) can also be built using the HAUV and the estimated Arrondissements' centroids computed from the Senegal map. The resulted averaged vector of the DHVs of each class shows a distribution of people referred to the target Arrondissement weighted by the distance displaced. This information has been expanded further by obtaining the "occupancy histogram" of each class. This histogram shows the number of people of the class that has occupied each Arrondissement along the time period comprised in the Data-set 3. This statistical characterization turns into a useful temporal characterization of people classes to be compared with other time series information, such as rainfall estimations, price changes, shocking events or seasonal cycles.

2.5.2. Validation of data processing

The resulting HAUV feature vectors have been compared to the vectors provided by a 1-step stationary Markov modelling of monthly displacements among Arrondissements. The correlation analysis between locations at different months derived from HAUV samples shows that they correspond to a non-stationary model: locations at summer months are less correlated with the rest of months locations; this validates seasonal (time dependent) population movements.

2.5.3. Rainfall estimations

Extracted from the TRMM-NASA project, they have been represented at different geographical resolution level in Senegal.

## 2.6. Visualization of mobility patterns and users' characteristic mobility profiles

Real decision making tools must provide detailed temporal and geographical resolution of the population movements. Therefore, three different web tools have been designed and implemented to visualize the variables of the proposed model in an integrated way, adjustable to different data dimensionalities and resolutions:

1. **Viz1** [10]: Visualization of daily series of variables (primary ones and variations) at the Arrondissement level. This visualization includes a map based representation of the variables as well as an Arrondissement correspondence graph to complement such representation.
2. **Viz2** [11]: Visualization of the resulting distribution of HAUVs for different groups of people (datasets, filtered populations, clusterized classes,…) by Arrondissements through time.
3. **Viz3** [12]: Visualization of the LZUVs also also for different types of population groups as a flow to understand the geographical distribution of the movements. It also embeds the visualization of the mobility profiles of the selected group and rain estimation diagrams.

## 3. RESULTS

The results obtained are:
1. Characterization of multi-scale mobility patterns for
    a. event detection;
    b. mobility profiling and calendarization of different communities.
2. We have characterized the relationship of mobility profiles with
    a. rainfall seasons;
    b. livelihood means;
    c. agricultural calendars.
3. For some regions, there are groups of inner population that show a yearly mobility profile in accordance to behaviors expected from other sources of information [3,7,8]. However, other groups display a profile which is not easily interpretable in such context and require further investigation. Even more, some regions seem to not have clear population groups following a specific pattern.

### 3.1. Multi-scale mobility patterns characterization

The different web tools developed allow for a characterization of mobility patterns at different aggregation levels with different applications.

### 3.1.1. Discovering events which drive strong mobility patterns abnormalities

**Viz1** [10] visualizes global movements among all Arrondissements: movements are coded via colors and arrows in a circle representing all the Senegal Arrondissements. For instance, daily user aggregated global movements can be represented, which is useful for general event detection.

Figures 2 to 4 show the potential of **Viz1** to understand and discover events or shock induced abnormalities in the mobility patterns. Figure 2 shows the behavior in a regular day: the color of each Arrondissement in the map based representation reflects the amount of mobility associated with it, whereas the Arrondissement correspondance graph provides a detailed origin-destination map of global movements (the color of each line represents the destination Arrondissement and its width is proportional to the amount of such movement; the size of each ring slice represents the total amount of people leaving such Arrondissement). Figures 3 and 4 illustrate the population movements corresponding to day numbers 355 and 357 of year 2013, when a national event occurred (Grand Magal at Touba). Therefore, the variations in the variables of the model may be exploited as a abnormality detection metric [5] to select candidates of significant events or shocks, when compared to the typical day characterization of movements in Senegal at a specific geographical level.

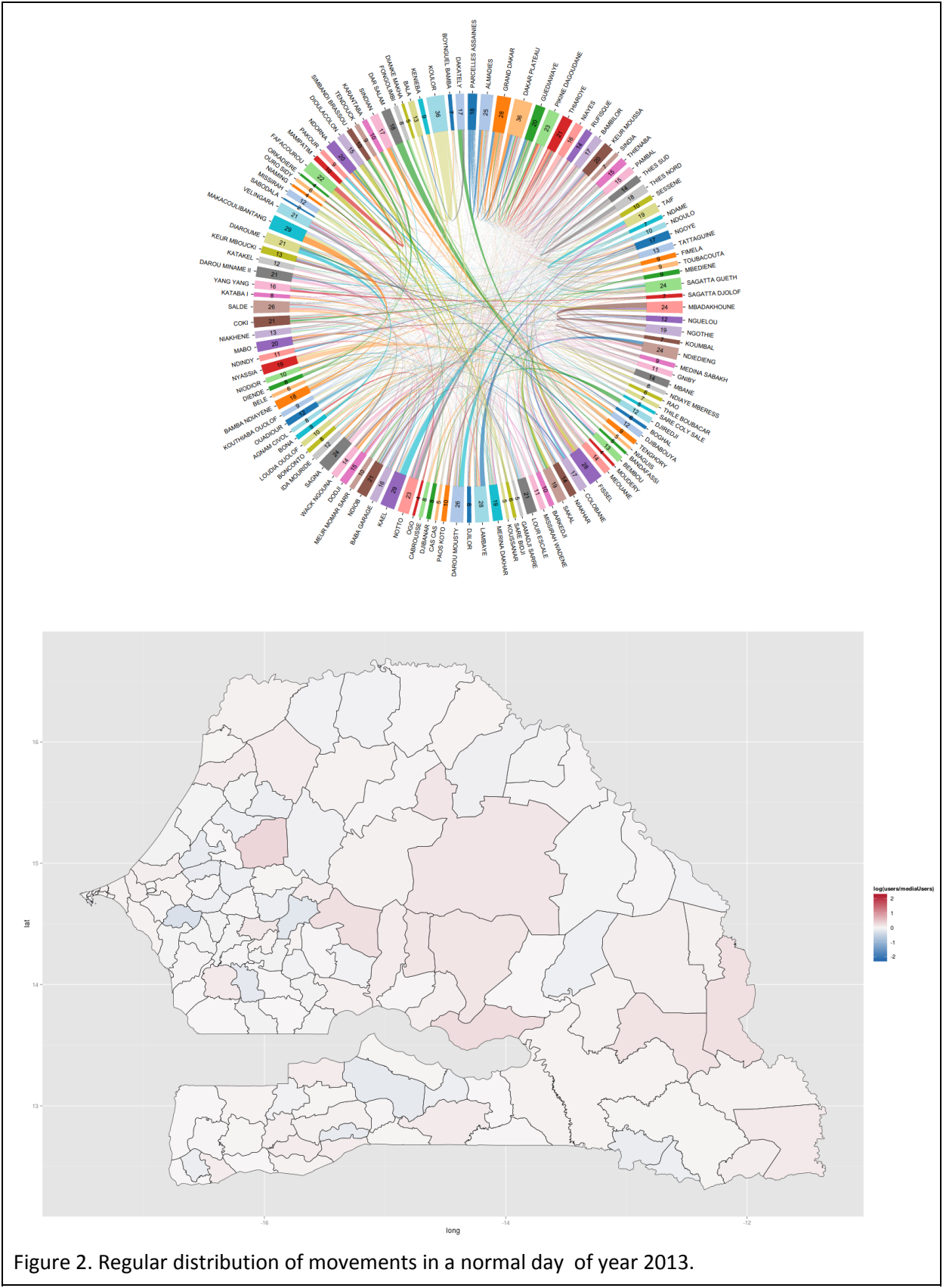

Figure 2. Regular distribution of movements in a normal day of year 2013.

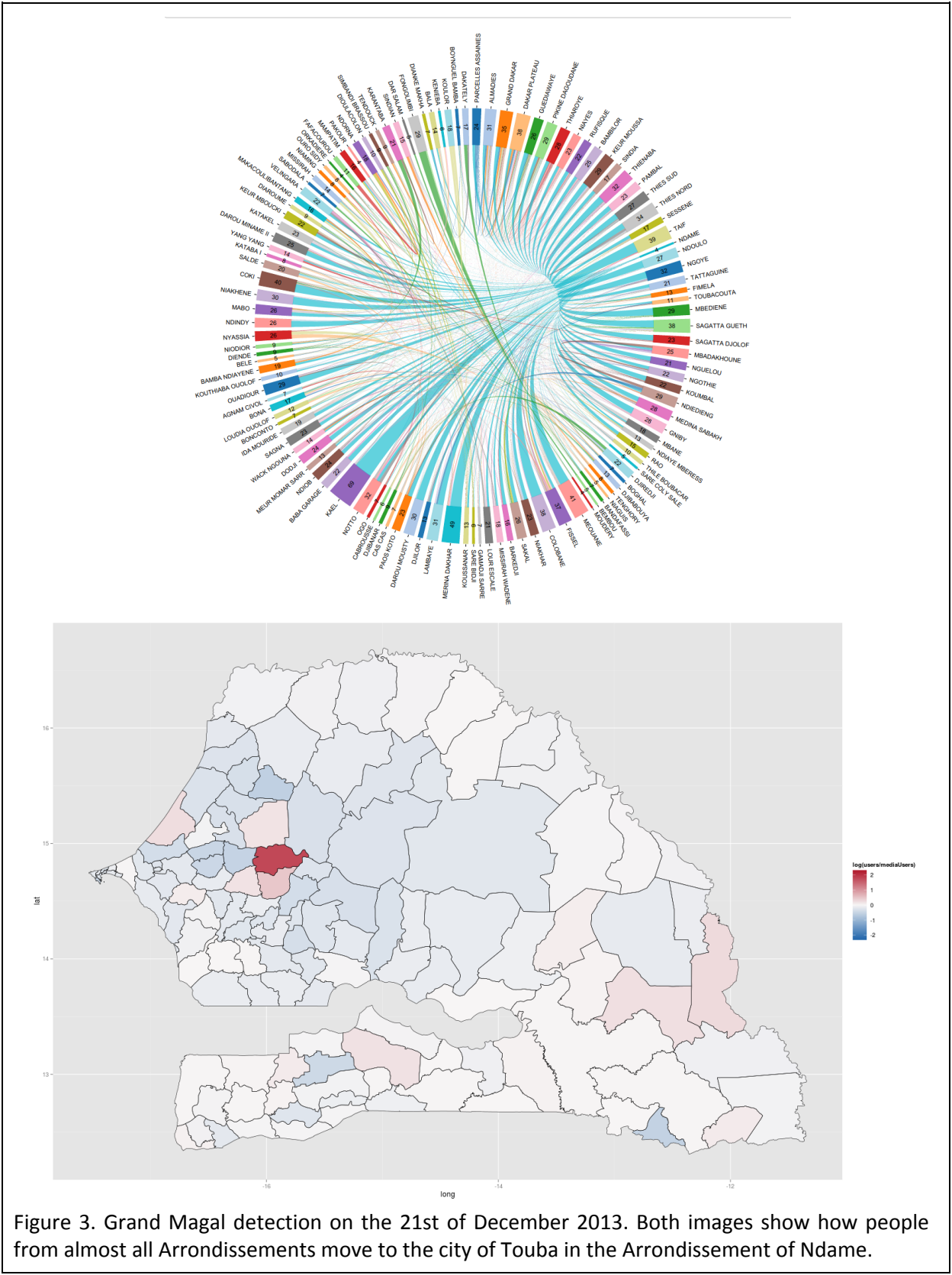

Figure 3. Grand Magal detection on the 21st of December 2013. Both images show how people from almost all Arrondissements move to the city of Touba in the Arrondissement of Ndame.

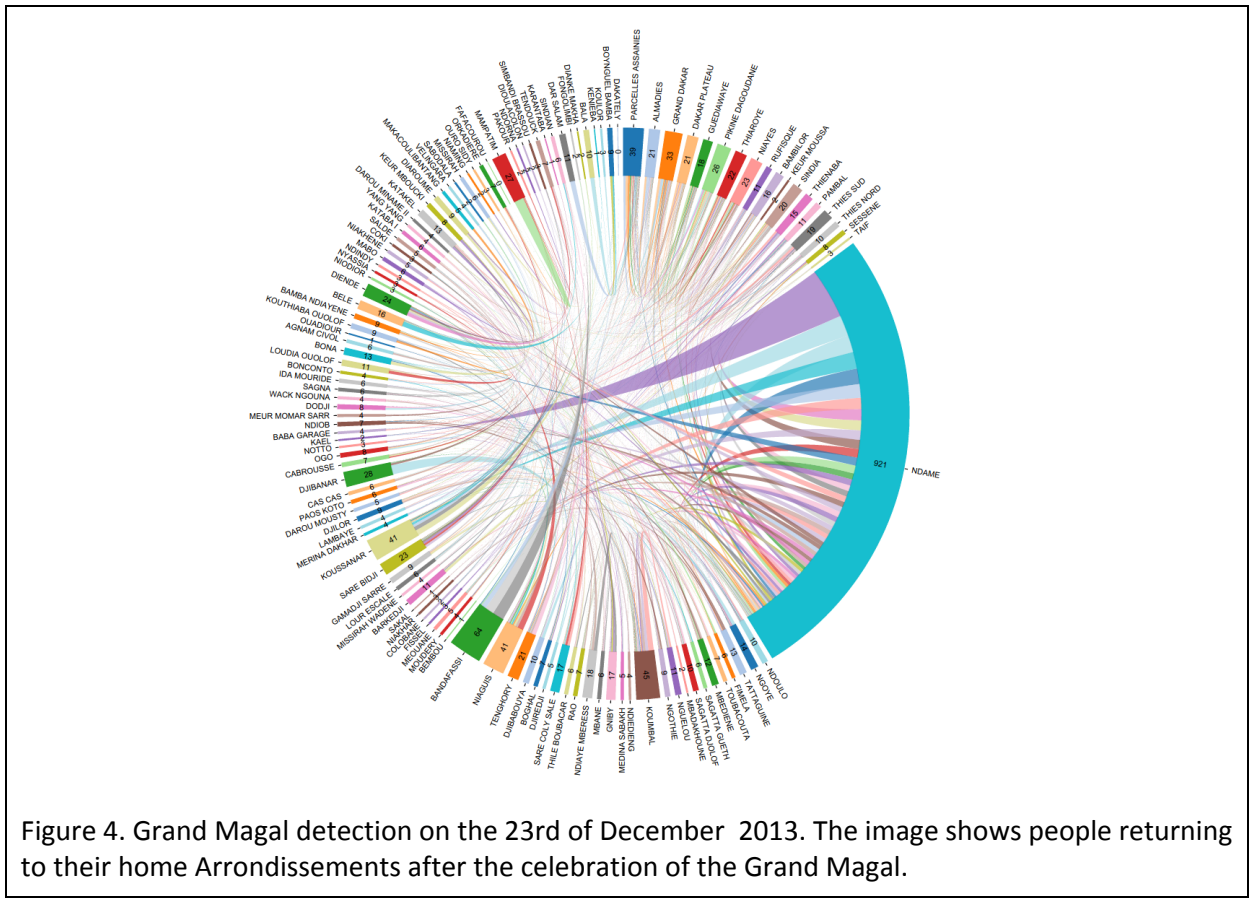

Figure 4. Grand Magal detection on the 23rd of December 2013. The image shows people returning to their home Arrondissements after the celebration of the Grand Magal.

### 3.1.2. Characterizing the destinations map of target populations through time.

**Viz2** [11] visualizes movements from/to a selected Arrondissement: the processed HAUVs (e.g., those corresponding to a class) are shown on Senegal´s map: for each month the number of moving people to each destination Arrondissement is color-coded. This is useful for low resolution movements.

Figure 5 shows how **Viz2** [11] helps to understand the mobility distribution of a population referred to one Arrondissement through the year.

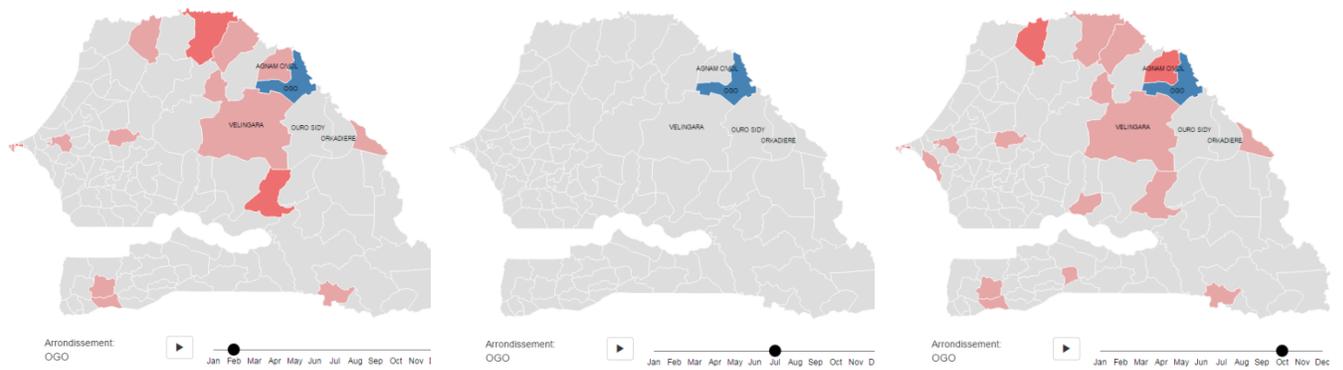

Figure 5: Distribution along Senegal for the population that was completely localized in the Arrondissement of MATAM during July.

This type of visualization shows how well organized the mobility profiles of specific population groups are; also, it helps to discover the preferential destinations of any mobility pattern and how this pattern evolves during the whole year.

### 3.1.3. Understanding flows of moving population between different livelihood regions

**Viz3** [12] visualizes movements from/to Livelihood Zones: for a given a set of HAUVs (selected by some of the developed classification techniques), the number of moving people to and from each destination livelihood zone is coded by size and color as well as indicated by the corresponding arrows on a Senegal map. This is useful for livelihood related movements.

The module allows the visualization of incoming (blue) and outcoming (red) flows of people for specific target populations selected (via any specified criterion) from the whole Senegal, depicting the interactions between the different livelihood zones. Here we illustrate the visualization of population groups provided by the clustering scheme explained in Section 2.3.2. By embedding the temporal profile of the target population and other external signals at the same scale, we can add more dimensions to these livelihood zones interactions. Figure 6 shows the mobility profiles corresponding to the group in Livelihood 8 (agropastoral zone specialized in peanut culturing) that occupies this zone towards the second half of the year after the rain season, and Figure 7 shows another group mobility profile in the same livelihood area corresponding to people who leave this zone during the rainy season.

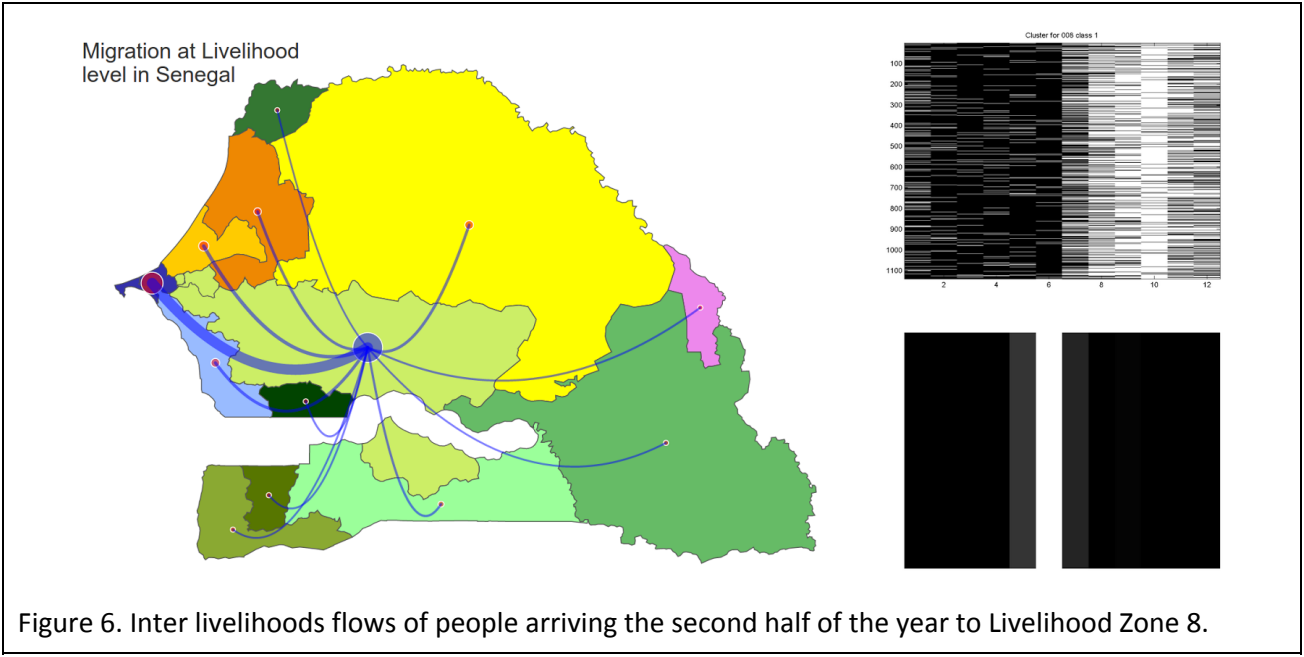

Figure 6. Inter livelihoods flows of people arriving the second half of the year to Livelihood Zone 8.

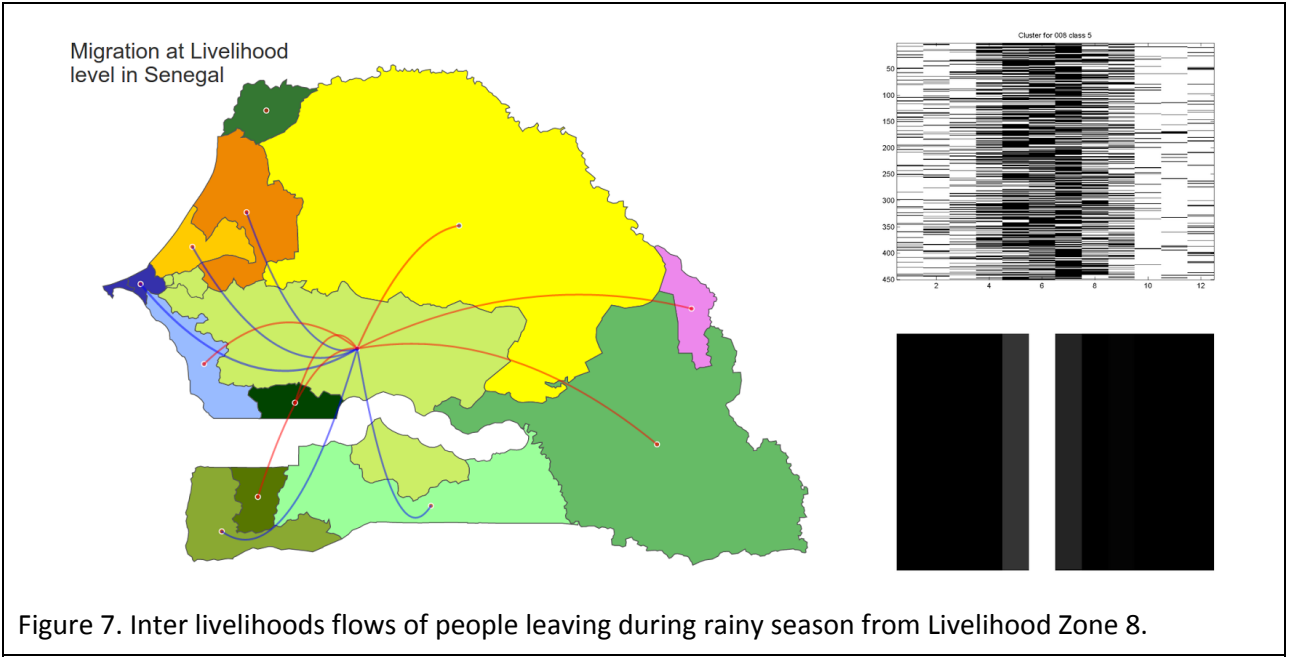

Figure 7. Inter livelihoods flows of people leaving during rainy season from Livelihood Zone 8.

## 3.2. Relationship between onset of large mobility changes and the end of rainy season

Population movements in the north of Senegal are expected to start in October due to the end of the rainy season [3]. A major potential advantage of the use of CDR data is that the onset of population movements may be estimated with high accuracy by measuring the changes in the mobility patterns of the users. This way, the actual reaction of the population to the change of season can be quantitatively measured.

3.2.1. Estimation of rainy season

Rainfall estimations have been extracted from [9][1]. The estimation has been calculated for Jan'11 until Dec'13, in order to observe yearly variations for a better interpretation. Fig. 8 summarizes the averaged rainfall by Arrondissements for the period observed; 2013 did not have significant changes when compared to 2011 (a less rainy year) or 2012.

.

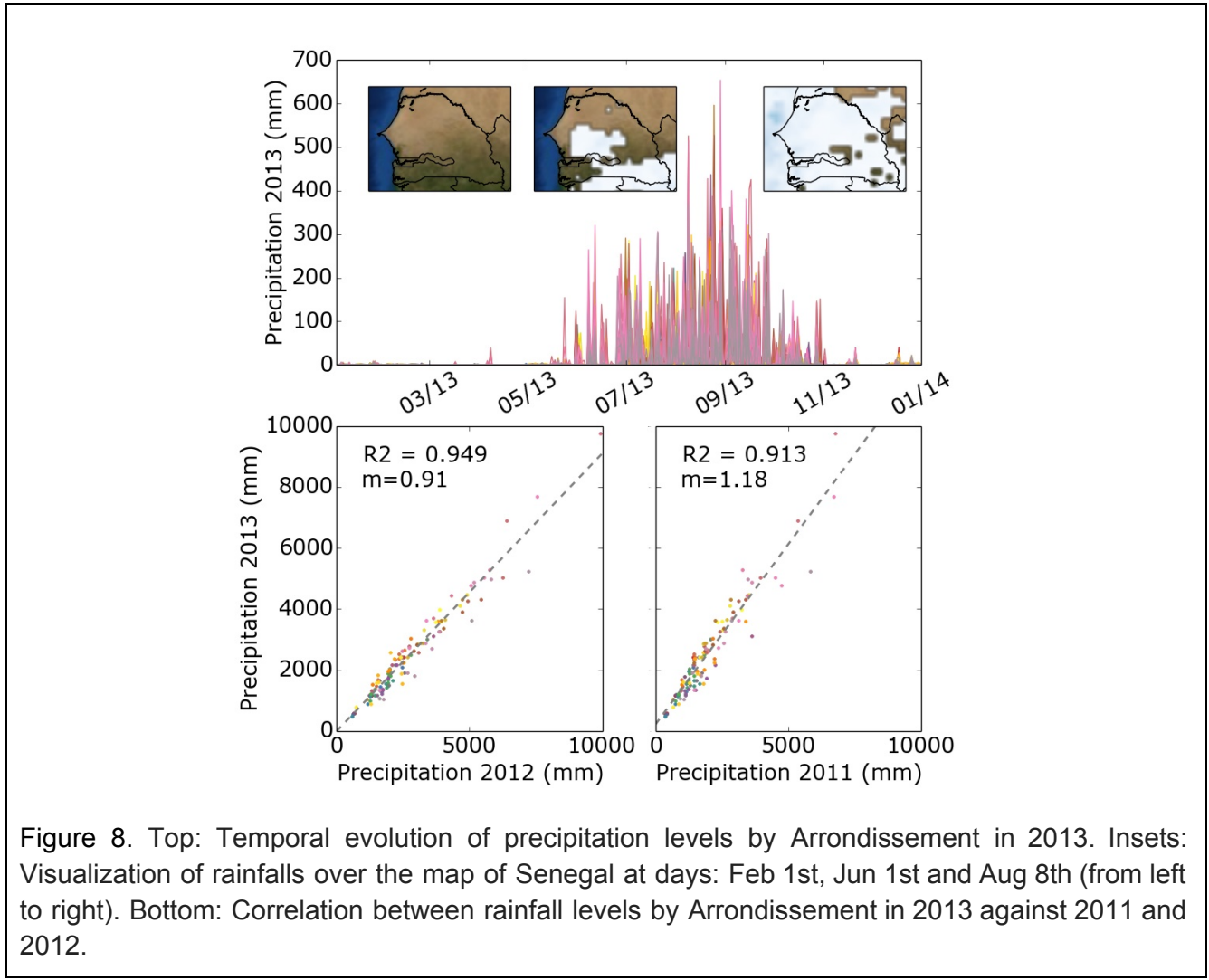

Figure 8. Top: Temporal evolution of precipitation levels by Arrondissement in 2013. Insets: Visualization of rainfalls over the map of Senegal at days: Feb 1st, Jun 1st and Aug 8th (from left to right). Bottom: Correlation between rainfall levels by Arrondissement in 2013 against 2011 and 2012.

3.2.2. Comparing the onset of mobility alterations with the end of rainy season

The results shown in Figures 6 and 7 feature the averaged rainfalls for the livelihood zone 8 (aggregated using the map in Fig. 1b) in a monthly time scale in order to simplify the analysis and visualization (as well as to compare them with agricultural calendars as explained below).

---
[1] http://erdos.mat.upm.es/d4d-senegal/rain.mp4

As observed, the selected mobility profiles are clearly influenced by the rainfall levels in the target candidate zone population that might be vulnerable to severe climate changes.

### 3.3. Discovering characteristic users' mobility profiles depending on the livelihood means

We have used both HAUVs and HLZUVs to obtain different classes (profiles) of temporal patterns of mobility regarding both Arrondissements and Livelihood Zones (LZs). Focusing on the latter, the clustering method (see materials and methods) provides groups of people that show the same occupancy profile in the target LZ; this classification can be cross-checked with the tagging of each LZ of Senegal according to the data in [2]. This process has been repeated for each of the Livelihood Zones in Senegal: some LZs provide expected results where other LZs show a non-easily interpretable behavior, requiring further consideration.

### 3.4. Visualizing correlation between mobility profiles and agricultural calendars

The characteristic profiles displayed in Figures 6 and 7 can also be time correlated with agricultural calendars (see [2]) as shown in Figures 9 to 11. The existence of correlation could spotlight groups of people that migrate depending on the agricultural cycles of livelihood zones of Senegal. Potentially, with a long-term characterization through several years, this strategy could help to monitor in real-time population vulnerable to climate changes or production alterations. However, detailed local analysis should be carried out to confirm and validate this hypothesis.

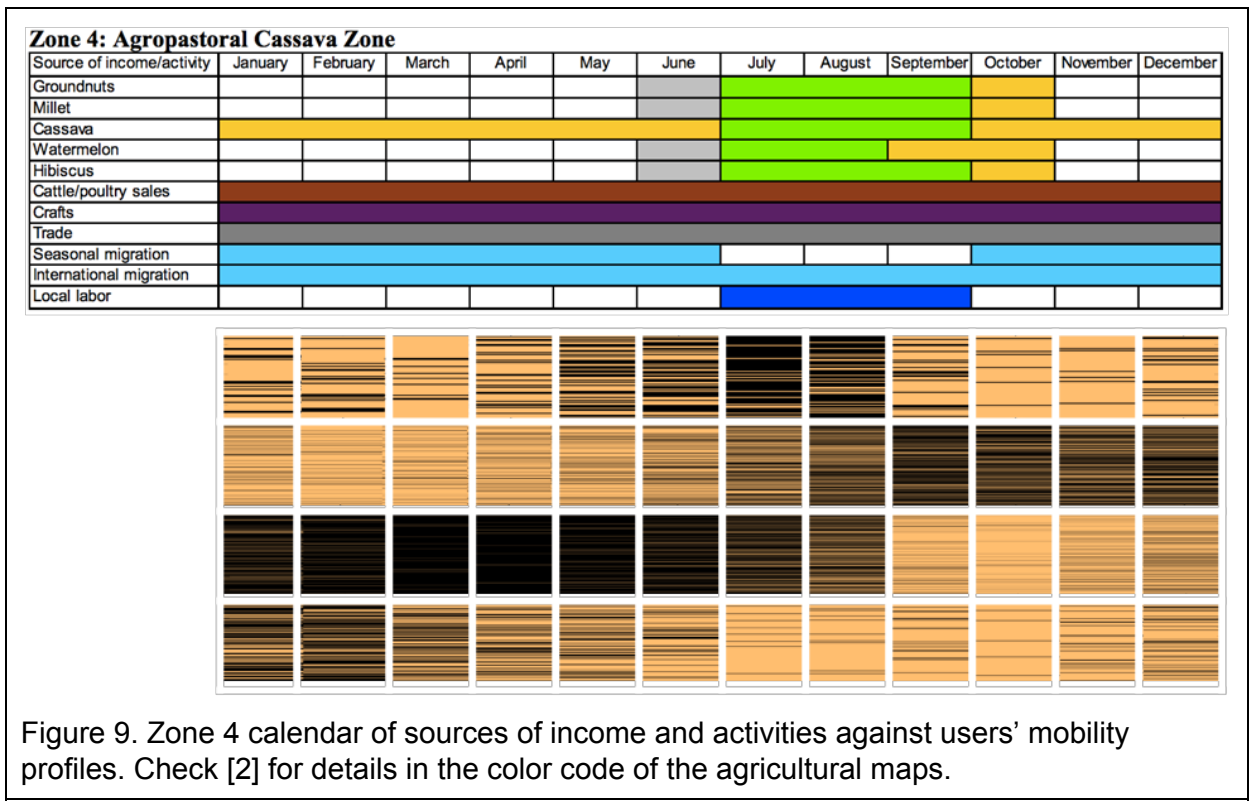

Figure 9. Zone 4 calendar of sources of income and activities against users' mobility profiles. Check [2] for details in the color code of the agricultural maps.

For instance, in Zone 4 (Fig. 9), the calendar interval for planting and weeding -green [2]- of resources (millet, cassava, watermelon, peanuts, hibiscus) which implies the rise of employment due to local labor -dark blue-, seems to trigger changes in several migration classes, although there is not a specific profile with a strong temporal correlation with this interval (only the third profile seems to correlate with a significant delay).

A more clear correlation between a mobility profile (row 1) and a calendar interval is found in the Zone 6 (Fig. 10) during the milk sales, as this region is specialized in herding and transhumant livestocks. As a hypothesis, variations in this correlation and the significance of this mobility profile could provide information about the success of the milk production and sales which would impact the regions economy and vulnerability.

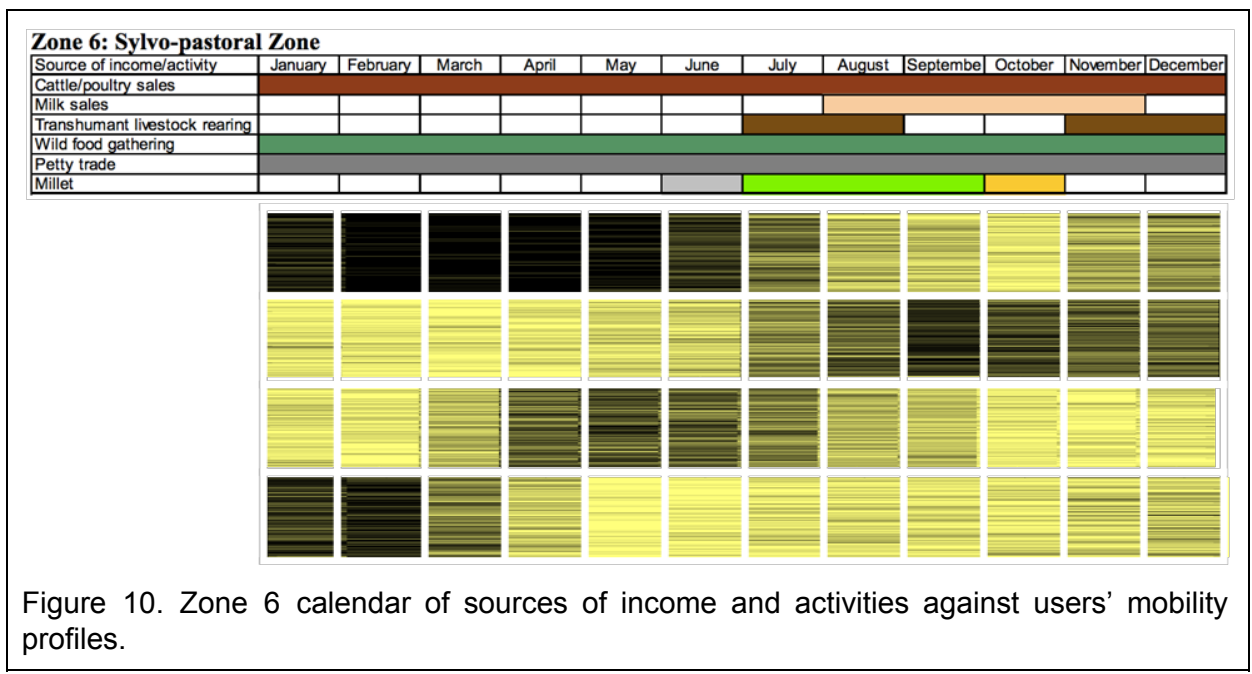

Figure 10. Zone 6 calendar of sources of income and activities against users' mobility profiles.

A very important livelihood zone of Senegal is the one specialized in peanut production (zone 8 in Fig. 11) in the central area of Senegal. The first two mobility profiles seem to correlate with the beginning of the planting preparation and the planting, when there is an increase of population due to the peanut production cycle. However, this is a very complex region involving seasonal migrations since it is a major transit zone from west to east and north to south; hence, further and more precise analysis is demanded to understand seasonal mobility through this zone. Another complex zone of study is the zone 13 (Fig. 12), where only the fourth mobility profile seems to correlate with the planting and collection of the crops of this zone.

It is important to notice that these calendars represent a normal characterization of the country production cycles. However, migrations could greatly change depending on external factors such as the rainfall levels, the market prices or extreme conditions or shocks. Only a long-term observation of calendars considering external variables (such as the estimated rainfalls) would enable to distinguish when the profiles are really driven by agricultural calendars and when they may be modulated by external factors.

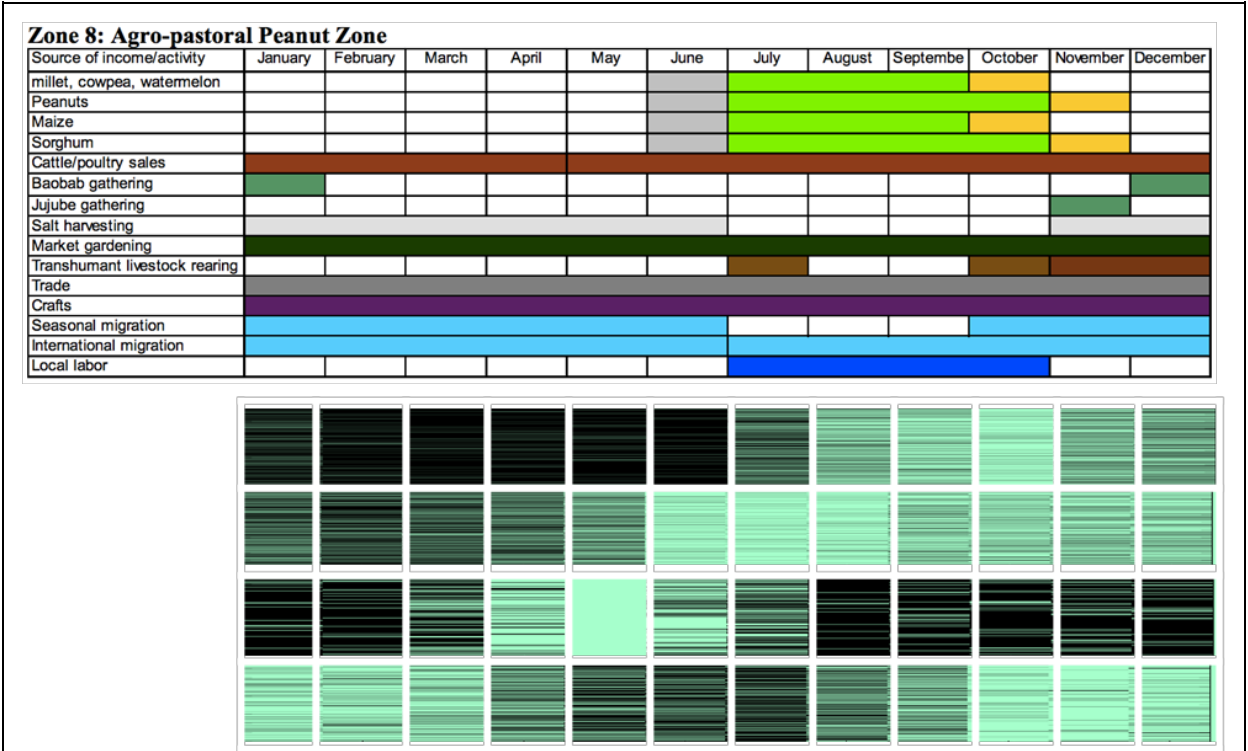

Figure 11. Zone 8 calendar of sources of income and activities against users' mobility profiles.

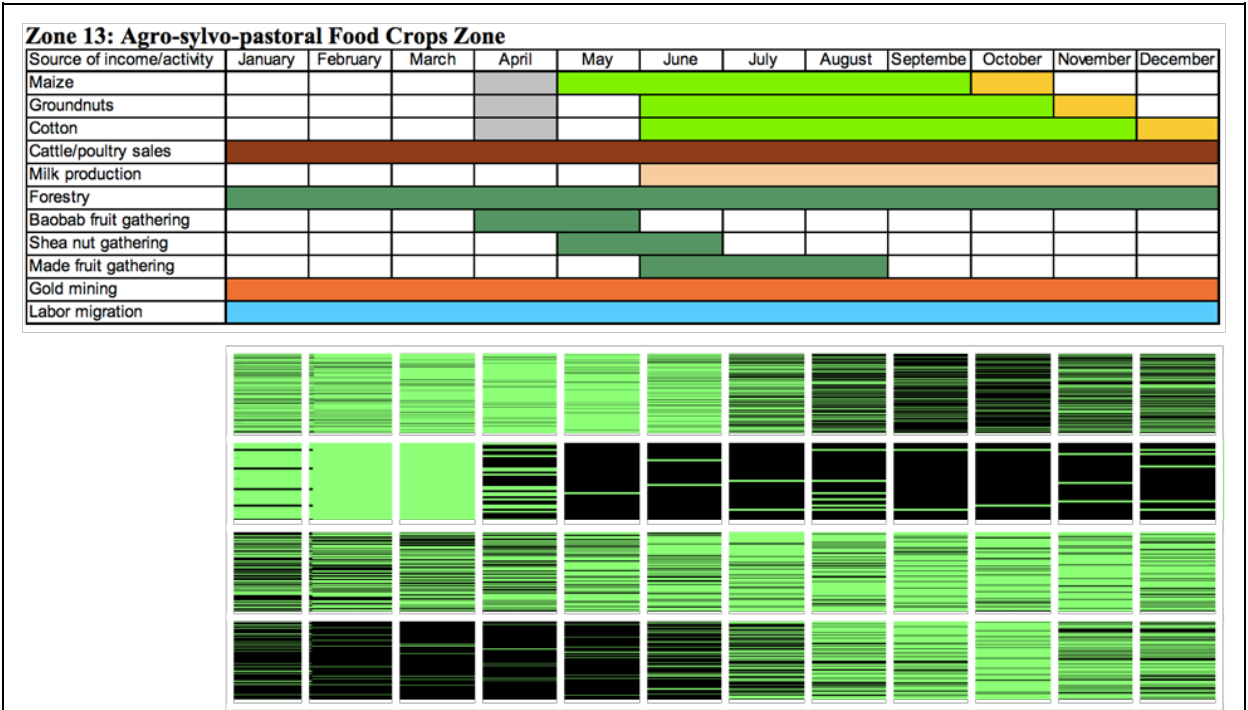

Fig. 12. Zone 13 calendar of sources of income and activities against users' mobility profiles.

## 4. DISCUSSION

This work has been motivated by the need of analyzing and quantifying the role of mobility patterns in the communities lifestyles and their access to basic resources, with the more precise and up-to-date information that the CDRs provide.

The developed processing and visualization prototype comprehensively integrates heterogeneous data for multiple use cases; here we have illustrated its potential by performing several off-line analyses (event detection, population mobility profiling and calendarization, etc.), with special focus on the possible interplay between mobility at the level of Livelihood Zones, accurate rainfall sensed data and agricultural calendars.

The time range limitation of the available Data-sets (an overall single year) does not allow for the robust design of on-line detection schemes, since seasonal reference baseline behaviors cannot be constructed. Nevertheless, the developed schemes can be easily extended to perform on-line detection provided larger time range data are available.

If the anonymization preserving additional limitations, either in time range or geographical resolution, of the Data-sets (fortnight time range for Data-set 2, Arrondissement resolution for Data-set 3) were removed, more robust and accurate results could be obtained. In general, this new approach to mobility patterns analysis could be very helpful to monitor vulnerable communities and to understand the impact of mobility patterns in the production means of Senegal.

## ACKNOWLEDGMENTS

The authors want to thank the support of Ministerio de Ciencia e Innovación of Spain via project MTM2010-15102, and Cátedra Orange at the ETSI Telecomunicación in the Universidad Politécnica de Madrid (UPM), Spain. They also want to thank Juan Fernando Sánchez-Rada for his help in developing some of the web applications.